\documentclass[reprint,amsmath,amssymb,superscriptaddress,aps,prl,twocolumn,showpacs]{revtex4}

\usepackage{standalone}
\usepackage{graphicx}
\usepackage{amsmath,latexsym}
\usepackage{sidecap}
\usepackage{color}
\usepackage{dcolumn}

\def\px{$p_x$}

\def\done{$d_1$}
\def\dtwo{$d_2$}
\def\dthr{$d_3$}

\def\QAg{Q_{\rm Ag}}
\def\QM{Q_{\rm Y}}

\begin{document}

\title{Electronic Instability and Anharmonicity in SnSe}

\author{Jiawang Hong}
\email{hongj@ornl.gov}
\affiliation{Materials Science and Technology Division, Oak Ridge National
Laboratory, Oak Ridge, TN 37831}

\author{ Olivier Delaire}
\email{olivier.delaire@duke.edu}
\affiliation{Materials Science and Technology Division, Oak Ridge National
Laboratory, Oak Ridge, TN 37831}
\affiliation{Mechanical Engineering and Materials Science Department, Duke University, Durham, NC 27708 }


\begin{abstract}
The binary compound SnSe exhibits record high thermoelectric performance,
largely because of its very low thermal conductivity. The origin of the strong
phonon anharmonicity leading to the low thermal conductivity of SnSe is
investigated through first-principles calculations of the electronic structure
and phonons.  It is shown that a Jahn-Teller instability of the electronic
structure is responsible for the high-temperature lattice distortion between the
Cmcm and Pnma phases. The coupling of phonon modes and the phase transition
mechanism are elucidated, emphasizing the connection with hybrid improper
ferroelectrics.  This coupled instability of electronic orbitals and lattice
dynamics is the origin of the strong anharmonicity causing the ultralow thermal
conductivity in SnSe. Exploiting such bonding instabilities to generate strong
anharmonicity may provide a new rational to design efficient thermoelectric
materials.
\end{abstract}


\maketitle


Quasiparticle couplings such as phonon-phonon and electron-phonon interactions
play a central role in condensed matter physics. In energy materials, it is critical to understand the interaction 
between phonons and the electronic structure, and how the chemical bonding impacts atomic vibrations, in order to
control both thermodynamics and transport properties. 
The origin of the strongly anharmonic bonding in binary metal chalcogenides is
the subject of renewed interest, both fundamentally and in connection with ferroelectric and
thermoelectric properties.  SnSe in particular 
is currently attracting strong interest, owing to the recent discovery of its
record-high thermoelectric efficiency.~\cite{zhao2014, zhao2016} This high
thermoelectric conversion efficiency results in part from an ultra-low thermal
conductivity.~\cite{zhao2014,chen_2014,carrete_2014,sassi_2014} 

Our recent inelastic neutron scattering (INS) measurements have shown that the ultralow
thermal conductivity of SnSe arises from strongly anharmonic phonons near a lattice instability.~\cite{chen_2015} 
A continuous structural phase transition occurs at $T_{\rm c}\sim805$\,K, associated to
the condensation of a soft phonon mode, which remains strongly anharmonic over a
broad temperature range below the transition. The soft mode is the lowest-energy
transverse optical (TO) phonon mode at the zone center in the low-symmetry phase
($T<T_{\rm c}$), but stems from a zone-boundary mode in the high-symmetry phase,
leading to a doubling of the unit cell on cooling through  $T_{\rm c}$.~\cite{chen_2015,
adouby_1998} The low-symmetry phase is non-polar. 

Thus, the behavior of SnSe is reminiscent of, while significantly
different from, the ferroelectric instability in the binary rocksalt
chalcogenides (PbTe, SnTe, GeTe). Upon cooling, SnTe and GeTe undergo a
displacive distortion from the cubic phase to a ferroelectric rhombohedral
structure, corresponding to the condensation of the degenerate transverse-optic
branch at the zone center ($\Gamma$). \cite{Jantsch1983} In the incipient
ferroelectric PbTe, the TO mode softening is only partial and the rocksalt phase
remains stable. As previously investigated, the soft-mode suppresses the thermal conductivity of rocksalt chalcogenides by scattering the acoustic phonons through anharmonic phonon-phonon interactions, which is beneficial to improve the thermoelectric
performance. \cite{Delaire_2011, Zhang_2011, Shiga_2012, Li_PRB2014, Li_PRL2014,
YueChen_PRL2014, Lee_2014}

Resonant bonding and lone-pair electrons have both
been associated with the lattice instability in the rocksalt compounds, as well
as in related phase-change materials. \cite{waghmare_2003, Lee_2014,
Wuttig_2008a} The compound SnSe is structurally quite distinct from the
rocksalts, however, with a layered structure giving rise to strongly anisotropic
electronic structure, lattice dynamics, and electrical and thermal transport
properties. It can be viewed as intermediate between the rocksalt structure and
the quasi-two-dimensional transition metal dichalcogenides, such as MoS$_2$ and
NbSe$_2$.

\begin{figure}
\includegraphics[width=2.5in]{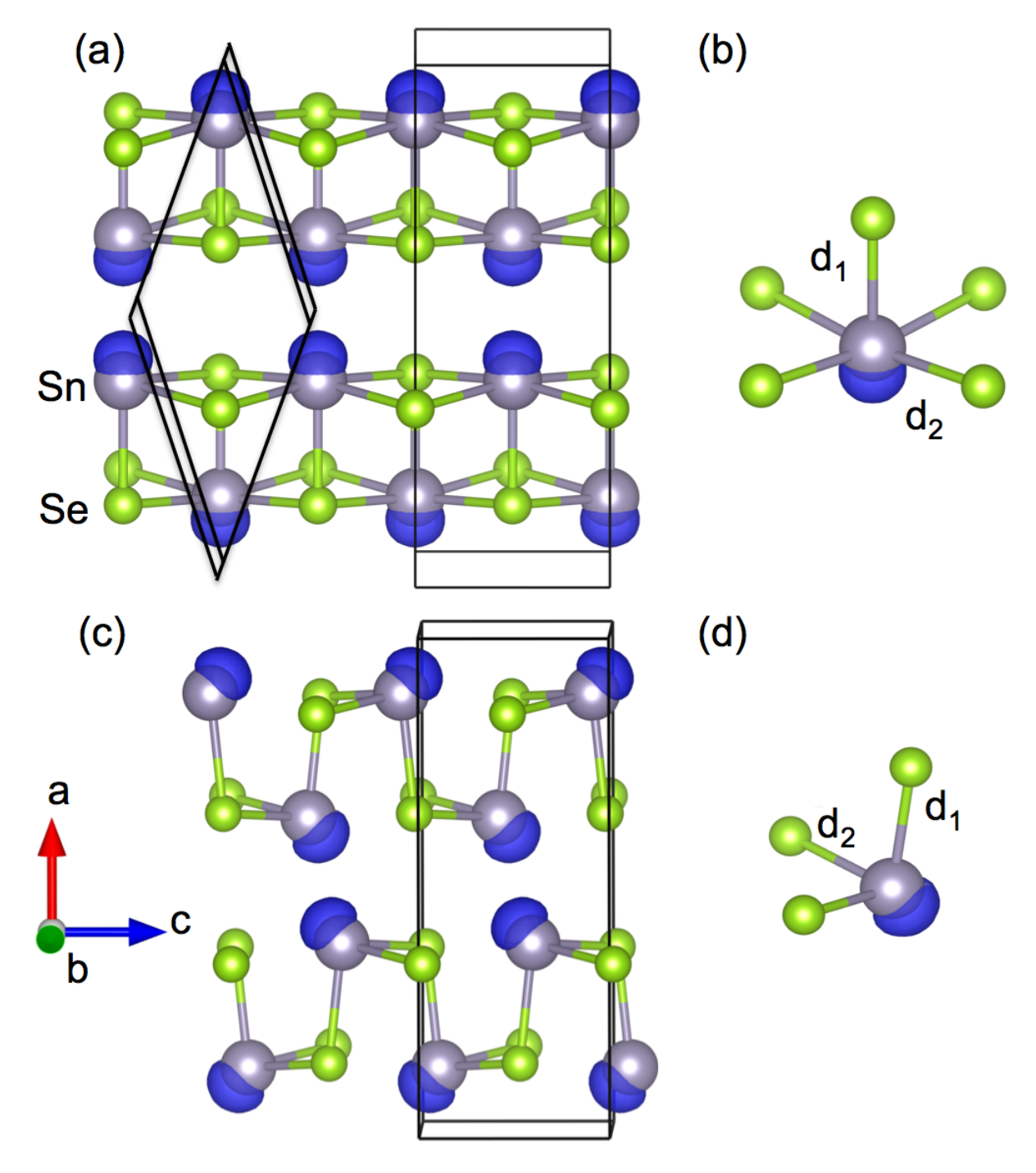}
\caption{\label{fig:structure}
(Color online) Crystal structure of SnSe in Cmcm (a) and Pnma (c) phases,
illustrating the double bilayer structure and Pnma distortion, and corresponding
Sn-Se bonding state (b,d). Sn atoms are in grey and Se atoms in green.  $d_1$
and $d_2$ are labels for bonds out-of-plane and in-plane, respectively.  The
crystal axes in the Cmcm structure are   chosen to match the Pnma phase, in
order to facilitate the comparison.  The conventional unit cell is indicated by
the black box, while the diamond shape indicates the primitive cell for Cmcm.
The blue caps on Sn atoms represent the isosurface of electron localization
function (ELF=0.93), showing the lone pair electrons of Sn atoms. }
\end{figure}

\begin{figure*}
\includegraphics[width=5in]{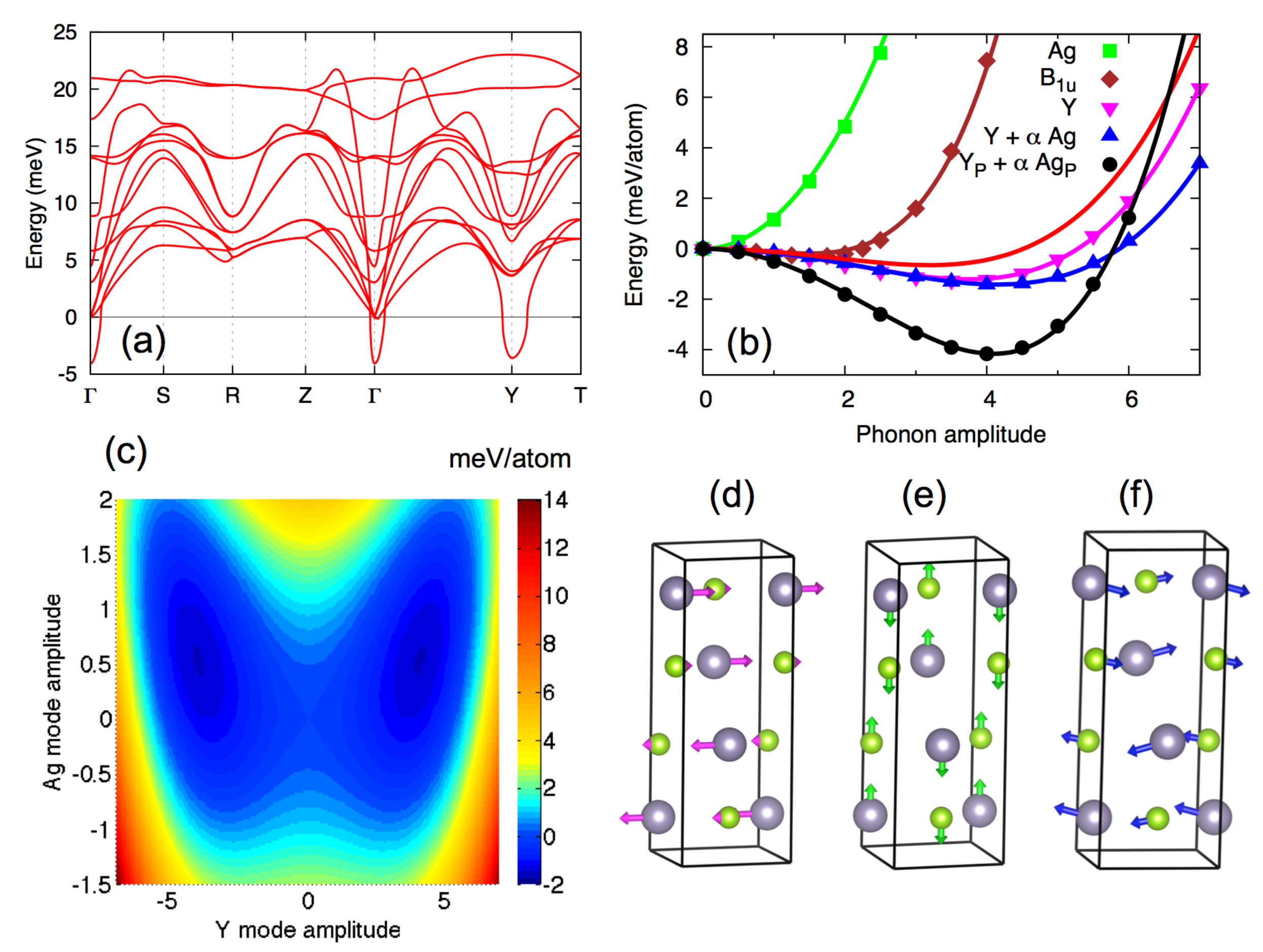}
\caption{\label{fig:dispersion_Cmcm}
(Color online) Phonon dispersions in Cmcm phase (a) and frozen-phonon potentials
(b) for the soft modes at  zone boundary ($Y$ mode) and the zone center
($B_{1u}$ mode and $A_g$ mode), and the combined modes of $Y + \alpha A_g$ in
Cmcm phase and pseudo Pnma phase ($Y_{\rm P} + \alpha A_{g \rm P}$), in which
the energy without distortion was shifted to the same energy as undistorted Cmcm
phase (set as 0) for clear comparison.  The symbols are {\it ab initio} energies
and the curves are fitted from Landau theory according to Eq.(1-3). The red
curve is a fit without the anharmonic coupling between $A_g$ and $Y$. (c) is the
frozen phonon energy for different coupling between $Y$ and $A_g$ mode. The unit
for the right color scale is meV/atom.  (d),(e) illustrate the phonon eigenvectors
for $Y$, $A_g$ modes in Cmcm phase, (f) is for the $A_g$ soft mode in Pnma
phase.  See text for more details.
}
\end{figure*}

SnSe crystallizes in a layered orthorhombic structure (Pnma) at ambient
temperature, with two bilayers along the $a$ direction. Upon heating, it
transforms continuously to the higher symmetry Cmcm phase at $T_{\rm
c}\sim$800\,K, ~\cite{Chatto1986,adouby_1998} as illustrated in
Fig.~\ref{fig:structure}. This second-order phase transition corresponds to the
condensation of the TO soft phonon mode of $A_g$ symmetry at the Brillouin zone
center ($\Gamma$) in the Pnma phase (see Fig.~\ref{fig:dispersion_Cmcm}f), as
predicted based on symmetry arguments, \cite{adouby_1998} and experimentally
observed with INS.\cite{chen_2015} However, a direct connection between
electronic structure and the strong anharmonicity in SnSe, which underpins its
ultra-low thermal conductivity, has remained elusive. This question is deeply
connected with the phonon mode instability and the microscopic mechanism of the
phase transition.  In this Letter, we show how the lattice instability in SnSe
arises from a Jahn-Teller-like electronic instability in the Cmcm phase, which
results in very large anharmonicity and ultra-low thermal conductivity.  In
addition, we investigate the anharmonically coupled zone-boundary and
zone-center phonon modes, which drive the phase transition from Cmcm to Pnma on
cooling across $T_c$.

The Pnma and Cmcm structures can be viewed as variants of the rocksalt
structure. In both phases, the Sn-Se bilayer is a rocksalt fragment, with the
two bilayers offset along the  $\langle 110 \rangle_{\rm cubic}$ direction, compared to rocksalt. The
Pnma phase is further distorted by an off-centering of Sn atoms in their
coordination polyhedra. Related distorted structures also appear in the MX
family (M=Ge, Sn; X=S, Se, Te), while PbX always crystallizes as rock salt
 at ambient pressure.~\cite{notebook}  In early investigations, Tremel and Hoffmann~\cite{Tremel1987} rationalized the
chemical bonding in SnS based on tight-binding electronic structure calculations
and showed that, within their 2-dimensional approximation, the mixing of the
conduction and valence bands play an important role in SnS. Waghmare {\it et
al.} focused on the stereochemical activity of cation lone pair electrons in
connection with the structural distortion in several MX compounds.~\cite{waghmare_2003} Here, we
reveal the origin of strong anharmonicity in SnSe as a coupling of specific phonons with the bonding instability, using accurate first-principles simulations.

%
Computations were performed in the framework of density functional theory (DFT)
as implemented in the Vienna Ab initio simulation package (VASP)
\cite{kresse_1993, kresse_1996}. All calculations used a plane-wave cutoff of
500 eV. We used the local-density approximation (LDA) and
projector-augmented-wave (PAW) potentials, explicitly including 4 valence
electrons for Sn (5$s2$ 5$p2$) and 6 for Se (4$s2$ 4$p4$). Our calculations used
LDA rather than the generalized gradient approximation (GGA) used in Refs.~\citenum{carrete_2014,skelton_2016}, since we previously observed that LDA phonon dispersions match better with INS
measurements.\cite{chen_2015} We used experimental structures \cite{adouby_1998}
as starting configurations and relaxed the lattice parameters and atomic
positions until all atomic force components were smaller than 1 meV/\AA.  The
phonon dispersions were calculated with VASP and Phonopy \cite{Togo_2008}, using
3$\times$5$\times$5 supercells (larger in-plan size is tested,
see Fig.S1). Based on convergence studies, we used
6$\times$12$\times$12 and 2$\times$4$\times$4 Monkhorst-Pack electronic
$k$-point meshes for the unit cell and supercell, respectively.  We used the
software Lobster to compute the crystal orbital hamiltonian populations
(COHP).~\cite{dronskowski_1993,deringer_2011, maintz_2013}

{\it Lattice instability.}---
The phonon dispersions (see Fig.~\ref{fig:dispersion_Cmcm}a) in Cmcm, in the
harmonic approximation, show soft modes at $\Gamma$ ($B_{1u}$ mode) and the zone
boundary $Y$ point (referred to as $Y$ mode thereafter~\cite{notes2}).
Similar phonon dispersion is reported in
	Ref.~\onlinecite{skelton_2016}.   At $\Gamma$,
the soft $B_{1u}$ mode is ``ferroelectric-like'': all Sn atoms move toward $+b$,
while all Se atoms move toward $-b$. In the $Y$ mode, Sn atoms in one bilayer move along $+c$ 
and Sn atoms in the other bilayer move along $-c$ ($b$ and $c$ are in Pnma coordinates here, see
Fig.~\ref{fig:dispersion_Cmcm}d). We describe in details below how this zone-boundary $Y$ mode overlaps
with the structural distortion to Pnma, and couples to a Jahn-Teller-like electronic instability.  
The $Y$ point in Cmcm becomes $\Gamma$ in Pnma and this soft mode reappears as the $A_g$
transverse optical mode in Pnma (Ref.~\cite{chen_2015}).

The calculated frozen-phonon potentials for the soft $B_{1u}$ and $Y$ modes are shown in
Fig.~\ref{fig:dispersion_Cmcm}b. Both modes exhibit a characteristic
``double-well'' profile (the potentials are symmetric and only the
positive side is shown). However, we note that the $B_{1u}$ mode only has a very
shallow well ($\Delta E=0.3$\,meV/atom), while the $Y$ mode shows a deeper
double-well ($\Delta E=1.25$ meV/atom), although still small.  The potential
energy surface $f(\QM, Q_{B_{1u}})$ computed from DFT is shown in supplementary
Fig.S2. It reveals that these two modes compete with each other, since the
energy has minima on the $Y$ or $B_{1u}$ axes, but increases for combinations.
The $Y$ mode
exhibits a stronger instability and it is the primary order parameter for the
distortion along $c$ that produces the Pnma structure, as experimentally
observed. Therefore, in the following we will mostly focus on
$Y$ mode and its coupling with $A_g$ mode.

According to the phase relationship between the experimental Pnma and Cmcm
structures, the zone-boundary soft mode at $Y$ is not sufficient to generate the
distortion observed, as some motion of atoms along $a$ is also needed (normal to
layers). This additional component is provided by an $A_g$ mode at $\Gamma$ with
energy 8.9\,meV (see Fig.~\ref{fig:dispersion_Cmcm}e).  We investigate the
coupling of $A_g$ and $Y$ modes, by expanding the potential energy in phonon
normal coordinates $\mathcal{Q}_n$ for phonon mode $n$ ($n=Y, B_{1u},
Ag$):
\begin{eqnarray}
f_n (Q_n) = a_n Q_n^2 + b_n Q_n^3 + c_n Q_n^4 \\
f_{\rm woc}(\QM, \QAg) =  f_{\rm Y}(\QM) + f_{\rm Ag}(\QAg)  \\
f_{\rm wc} (\QM, \QAg) =  f_{\rm Y}(\QM) + f_{\rm Ag}(\QAg)  \nonumber \\
+ d \QAg \QM^2  + e \QM^2 \QAg^2 .
\end{eqnarray}
Here we have made the approximation of truncating the expansion systematically
at overall fourth order, and $\mathcal{Q}_n = ({Q_n}/\sqrt{N m_j}) {\rm
Re}[\mathbf{e}_j \, {\rm exp}(i \mathbf{q} \cdot \mathbf{r}_{jl})]$, $N$ is the
number of atoms in the supercell, $m_j$ is the mass of atom $j$, $\mathbf{q}$ is
the wavevector, $\mathbf{r}_{jl}$ is the position of atom $j$ in unit cell $l$,
and $\mathbf{e}_j$ is the $j$-th atom component of the eigenvector.  $Q_n$ is
the phonon amplitude of mode $n$, and $a_n, b_n, c_n, d,e$ are numerical
parameters used to fit DFT energies, with $b_n=0$ for $Y$ and $B_{1u}$ modes,
owing to mirror symmetry along $c$ and $b$ directions in Cmcm,
respectively. Here, $f_{\rm Y}(\QM)$ and $f_{\rm Ag}(\QAg)$ are the frozen
phonon potentials along individual coordinates, while $f(\QM, \QAg)$ corresponds
to the superposition of these two modes. We attempt to express the latter both
without coupling, $f_{\rm woc}(\QM, \QAg)$ (Eq.(2)), and with mode coupling
through $d$ and $e$ terms in Eq.(3),~\cite{notes-symm} denoted
$f_{\rm wc} (\QM, \QAg)$. Such couplings of zone-center and
	zone-boundary modes are of strong current interest in improper hybrid
	ferroelectric oxides.~\cite{bousquet_2008, benedek_2011}

\begin{figure}
\includegraphics[width=3.0in]{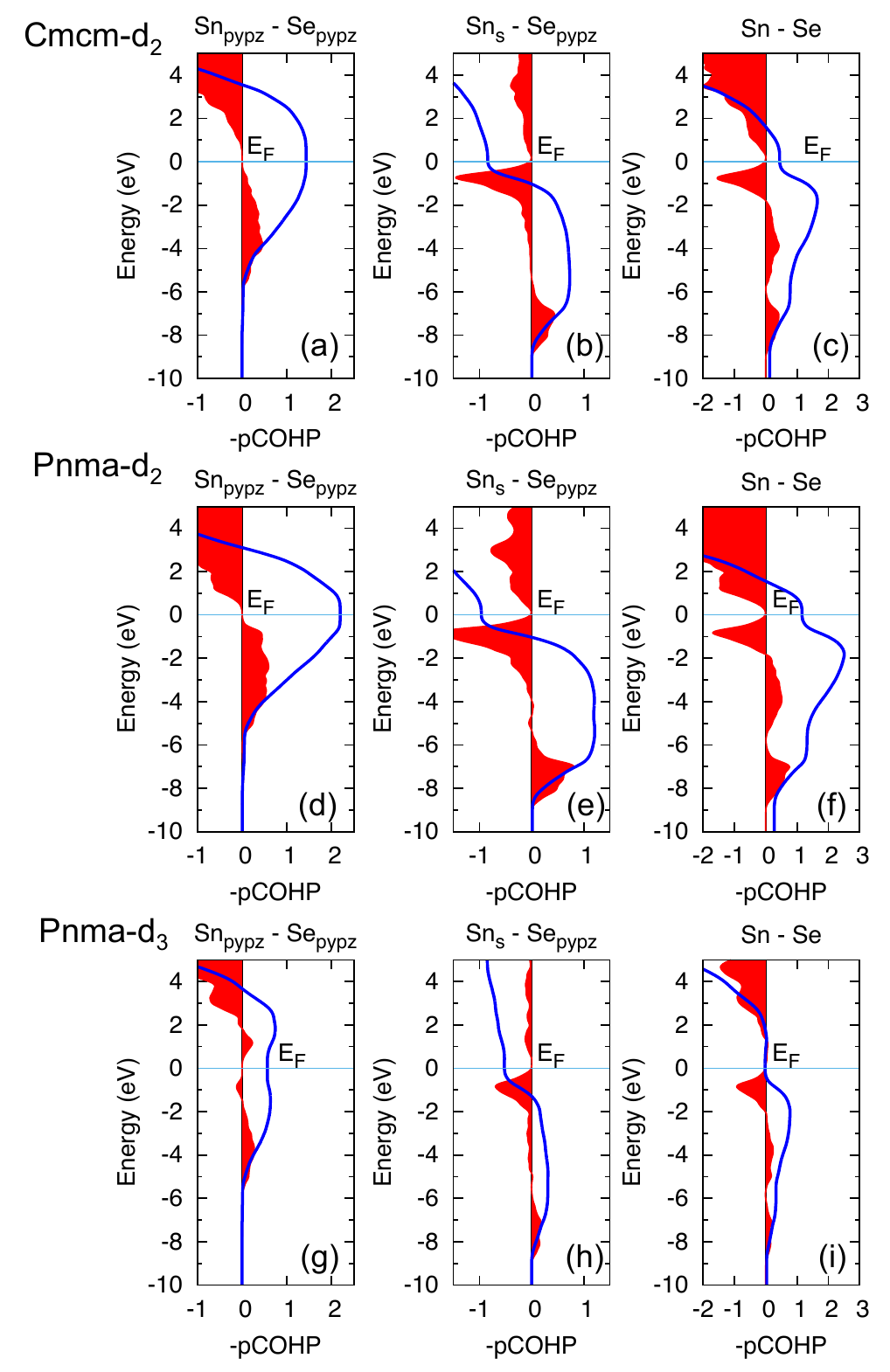}
\caption{\label{fig:COHP}
(Color online)  COHP calculation for Cmcm \dtwo\ bond (top panels) and Pnma
\dtwo\ bond (middle panels) and Pnma \dthr\ bond (bottom panels).  The first
column (a,d,g) is the bonding interaction between in-plane orbitals Sn-5$p$ and
Se-4$p$; the second column (b,e,h) is the bonding and anti-bonding interaction
between lone pairs Sn-5$s$ and in-plan Se-$p$ orbitals; the third column (c,f,i)
is the total bonding interaction between Sn and Se atoms. Red filled curves are
the $-$pCOHP and the blue curves are $-$IpCOHP.
}
\end{figure}

The potential energy surface $f(\QM, \QAg)$ computed from DFT is shown in
Fig.~\ref{fig:dispersion_Cmcm}c. As can be seen from this figure, the two modes
are clearly coupled, with a diagonal minimum for the energy at an amplitude ratio
$\alpha= \QAg / \QM =0.15$, close to the experimental value
of the structural distortion $\alpha \simeq 0.2$. The DFT energies
(markers) in Fig.~\ref{fig:dispersion_Cmcm}b are fit well using Eq.(1-3), shown
as lines on the same figure, indicating the model is appropriate. We note for
$d=e=0$ (i.e.  without coupling between $\QM$ and $\QAg$), the resulting energy
curve for $\QAg = 0.2\times \QM$  (red curve) is clearly higher than for pure
$Y$ mode. A satisfactory fit requires the anharmonic coupling terms in Eq.(3),
as shown by the blue curve. We also plot the frozen-phonon potential for the
$A_g$ mode alone and the linear combination with $\QAg = 0.2\times \QM$. This
shows that the potential for $A_g$ alone is stable and increases sharply with
phonon amplitude, while the combined mode reaches a lower energy than the pure
$Y$ mode. These results clearly establish the importance of anharmonic coupling
between $Y$ and $A_g$ modes in SnSe.

We note that even the combined $Y + A_g$ distortion lowers the energy by only
1.45\,meV/atom, however, much smaller than the total predicted energy difference
between relaxed theoretical Cmcm and Pnma phases ($\sim$5\,meV/atom), but this discrepancy can be attributed to the strain.  We
further explored the respective roles of atomic displacements and lattice strain
by computing the distortion potential ($\QAg + \alpha \QM$, $\alpha=0.2$) in a
pseudo-phase based on the optimized Pnma lattice constants but with the Cmcm
internal atomic positions. The result is shown in
Fig.~\ref{fig:dispersion_Cmcm}b (black markers and line).  The depth of the
double-well is now very close to the Cmcm-Pnma
energy difference, confirming the importance of strain.

{\it Electronic instability.}---
We now explain how the lattice instability and mode coupling result
from the underlying electronic instability.  We start by discussing the
projected crystal orbital hamiltonian populations (pCOHP). The
related approach of crystal orbital overlap population (COOP) analysis
provides electron-resolved bonding information. But we chose the COHP to
provide energy-resolved local bonding information, well suited to probe
the electronic instability.~\cite{dronskowski_1993,deringer_2011,
maintz_2013}
The pCOHP was calculated for bonds
$d_1$, $d_2$ (see notation in Fig.~\ref{fig:structure}), in both Cmcm and Pnma phases and
\dthr\ corresponding to the two elongated in-plane bonds in Pnma. 
The integrated
pCOHP (IpCOHP) is also calculated to probe the energetics of bonding and
antibonding interactions, as its magnitude correlates with the extent of
covalency and relative bonding strength (the more negative value,
the stronger covalent bonding strength). The total pCOHP and IpCOHP between the
neareast Sn and Se and the main contribution from in-plane Sn $p$ and Se $p$
orbitals, and the Sn $s$ and Se in-plane $p$ orbitals are shown in
Fig.~\ref{fig:COHP}.~\cite{notes-COHP} The IpCOHP ($\xi$), bond lengths ($l$) and force constants
($k$) for \done, \dtwo\, and \dthr\ are compared for both phases in
Table~\ref{tab:IpCOHP}. A schematic illustration and
explanation of negative force-constants in this table can be found in
Fig.S3.

\begin{table}
\caption{Bond length ($l$), IpCOHP ($\xi$) and force constant ($k$) for \done, \dtwo\
	and \dthr\ bonds in Cmcm and Pnma phases.  Total IpCOHP for Sn
	polyhedron (\done + 2\dtwo + 2\dthr) is also listed. The force constant
	is the average of in-plane component for \dtwo\ and \dthr\ ($k_{yy}$ +
	$k_{zz}$)/2 and axial component for \done\ ($k_{xx}$). $l$ in \AA,
	$\xi$ in eV, $k$ in eV/\AA$^2$.
	\label{tab:IpCOHP}
}
\begin{ruledtabular}
	\begin{tabular}{clrrrc}
	    &	& \done & \dtwo & \dthr & Total \\
		\hline
           &  $l$    &  2.71   &   2.96    & 2.96    &  --    \\ 
      Cmcm &  $\xi$  &$-$0.96  &  $-$0.45  & $-$0.45 & $-$2.76     \\ 
           & $k$     & $-$3.92 &  $-$0.35  & $-$0.35 &  --    \\ 
           &  $l$    & 2.74    &   2.79    &  3.20   &  --    \\ 
      Pnma &  $\xi$  & $-$0.75 & $-$1.16   & 0.02    & $-$3.03     \\
           & $k$     & $-$3.31 &  $-$1.32  &  0.01   &  --    \\ 

	\end{tabular}
\end{ruledtabular}
\end{table}

From Fig.~\ref{fig:COHP}, one can see that the most energetically favorable
interactions are in-plane ($yz$ plane) Sn 5$p$ and Se 4$p$.  The hybridization
between Sn-$s$ and in-plane Se-$p$ shows strong occupied anti-bonding states
just below the Fermi level. In Cmcm,  four degenerate Sn-Se bonds share four
electrons, forming identical ``resonant'' half-filled bonds along $\langle 011
\rangle$.  This high-symmetry resonant bonding state is energetically
unfavorable.  In distorted Pnma at low $T$, a geometric distortion breaks the
symmetry and causes two shorter (\dtwo) and two longer (\dthr) bonds. The
shorter \dtwo\ bonds enhance significantly the in-plane Sn-$p$ -- Se-$p$
interactions (from $\xi=-1.43$ to $\xi=-2.20$ eV, Fig.~\ref{fig:COHP}a,d), as
well as anti-bonding hybridization of Sn-$s$ and in-plane Se-$p$ to a lesser
degree (from $\xi=0.84$ to $\xi=0.96$ eV, Fig.~\ref{fig:COHP}b,e).  The \dtwo\
bond thus becomes much stiffer and the force-constant ($k^{d_2}_{yy}$ +
$k^{d_2}_{zz})/2$ increases nearly four-fold from $-0.35$\,eV/\AA$^2$ to
$-1.32$\,eV/\AA$^2$, as shown in Table~\ref{tab:IpCOHP}.  The IpCOHP for \dtwo\
bond decreases from $-0.45$ eV in Cmcm is to $-1.16$ eV in Pnma,
showing significant strengthening of \dtwo\ bonds.
Simultaneously, the \dthr\ bonds weaken to near-zero IpCOHP and
force-constant, indicating \dthr\ breaks in the Pnma phase, as the Sn
coordination changes to a triangular pyramid with almost equal \done\ and \dtwo\
bond lengths (see Fig.~\ref{fig:structure}d and Fig.S5).  This large difference between
\dtwo\ and \dthr\ causes the large non-linear forces (anharmonicity) for Sn motion along $c$, as the strongly
anharmonic frozen phonon potential for $Y$ and $Y+\alpha A_g$ modes shows in
Fig.~\ref{fig:dispersion_Cmcm}b.

In contrast, the \done\ bonding changes little between the two phases (Fig.S4).
On cooling, \done\ slightly elongates and its force-constant weakens a little,
compatible with the IpCOHP. Again, the Sn 5$s$ and Se 4\px\ have occupied
anti-bonding states just below the Fermi level ($\xi=0.49$\,eV). However, the Sn
5\px\ and Se 4\px\ states are strongly bonding ($\xi = -1.47$\,eV) and overall
the \done\ bond is stable and does not change much through the transition.  From
Cmcm to Pnma, the total IpCOHP (summing over the Sn nearest-neighbor bonds)
decreases from $-2.76$\,eV to $-3.03$\,eV (Table~\ref{tab:IpCOHP}), showing that
the latter is electronically more stable.  Therefore, the distortion is
electronically driven by lowering the electronic energy through lifting the
degeneracy of ``resonant'' bonding in the Cmcm phase, similar to a Jahn-Teller
distortion.~\cite{JT1937,pearson_1975}

Finally, we investigate the behavior of the Sn 5$s$ lone-pair electrons in the
phase transition.  In Fig.~\ref{fig:structure} and Fig.S5, the plots of electron
localization functions (ELF)~\cite{silvi1994} clearly show that the Sn 5$s$
lone-pair electrons are stereochemically active in both phases.  As discussed
above, the Sn 5$s$ lone-pair combines with the Se 4$p$ states to form a bonding
state at $\sim-8$ eV and an occupied anti-bonding state at the top of the
valence band.  We also observe an admixture of states across the gap (see
Fig.S6).  The orbital-weighted band structure (Fig.S6) and projected density of
states (Fig.S7) show that the bottom of the conduction band is mostly composed
of Sn 5$p$ bands, while the top of the valence band is mainly derived from Sn
5$s$ and Se 4$p$ anti-bonding hybrids, as also evident in Fig.~\ref{fig:COHP}.
Along the $\Gamma$-Y direction (Fig.S6), one can clearly see the mixture of
conduction band and valence bands, which is also responsible for the
stabilization of the lower symmetry Pnma, as previously suggested by Tremel and
Hoffmann.~\cite{Tremel1987}.

This can be understood as Sn 5$p$ states and anti-bonding  (Sn5$s-$Se4$p$)$^*$
states mixing into a new bonding ($m$) state and anti-bonding ($m^*$) state,
where $m$ is just below $E_{\rm F}$ and the anti-bonding state above $E_{\rm
F}$. This stabilization is achieved by distorting from the rocksalt structure
and mixing the valence and conduction bands \cite{Tremel1987}.  The orbital
stabilization requires asymmetric electron density where the lone pair
distribution is projected away from the cation (Sn$^{2+}$) toward the interlayer
void region. This is illustrated in Fig.~\ref{fig:structure}d and Fig.S5.

In summary, the lattice distortion in SnSe is driven by a Jahn-Teller electronic
instability, which results in strong anharmonicity of lattice dynamics. In
particular, zone-boundary and zone-center phonon modes are anharmonically
coupled, explaining the observed structural phase transition path from Cmcm to
Pnma on cooling across $T_c$. This behavior is similar to coupled phonon instabilities in hybrid improper ferroelectrics, although the low-$T$ phase of SnSe remains non-polar. The instability of the in-plane resonant bonding
in this quasi-two-dimensional structure is the chemical origin of the strong
anharmonicity, and drives the Jahn-Teller distortion. This mechanism could provide a way to design
materials with low thermal conductivity, suitable for the thermoelectric
applications, by tuning the chemistry and orbital interactions. 
 

\begin{acknowledgments}
This work was supported by the U.S. Department of Energy, Office of Science,
Basic Energy Sciences, Materials Sciences and Engineering Division, through the
Office of Science Early Career Research Program (PI Delaire). This research used
resources of the Oak Ridge Leadership Computing Facility, which is supported by
the Office of Science of the U.S. DOE. 
\end{acknowledgments}


\bibliography{snse}

\begin{thebibliography}{35}
\bibitem[{not({\natexlab{b}})}]{notes2}
\bibinfo{note}{In original Cmcm phase, $Y$ point is (0.5,0.5,0), this $Y$ point
  is (0.5,0,0.5) in our notation.}
\bibitem[{not({\natexlab{c}})}]{notes-symm}
\bibinfo{note}{The coupling term $Q^2_{Ag} Q_Y$ is omitted because the energy
  is an even function for $Y$ mode, due to mirror symmetry along $c$ diction in
  the Cmcm phase.}
\bibitem[{not({\natexlab{d}})}]{notes-COHP}
\bibinfo{note}{We show the negative values ($-$COHP and $-$IpCOHP) in the Fig.3
  to resemble the normal COOP plots, i.e. drawing bonding contributions on the
  right of vertical line of pCHOP=0, and antibonding contribution on the left.}
\end{thebibliography}

\cleardoublepage
\setcounter{figure}{0}
\renewcommand\thefigure{S\arabic{figure}}


\begin{center}
	{\large {\textbf{Support Information: Electronic Instability and Anharmonicity in
	SnSe}}}
\end{center}


\begin{figure*}
\includegraphics[width=4.5in]{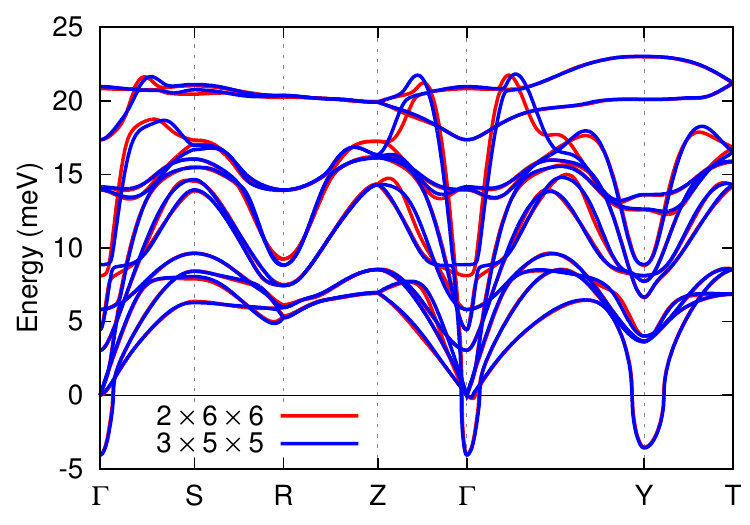}
\caption{\label{fig:supercell_size}
(Color online)  Convergence test for supercell size in-plane. The almost
overlapped phonon dispersion curves between two supercell sizes in-plane ($5\times5$ vs.
$6\times6$) suggest the in-plane size chosen in our calculation
is large enough.
} 
\end{figure*}

\begin{figure*}[t]
\includegraphics[width=4in]{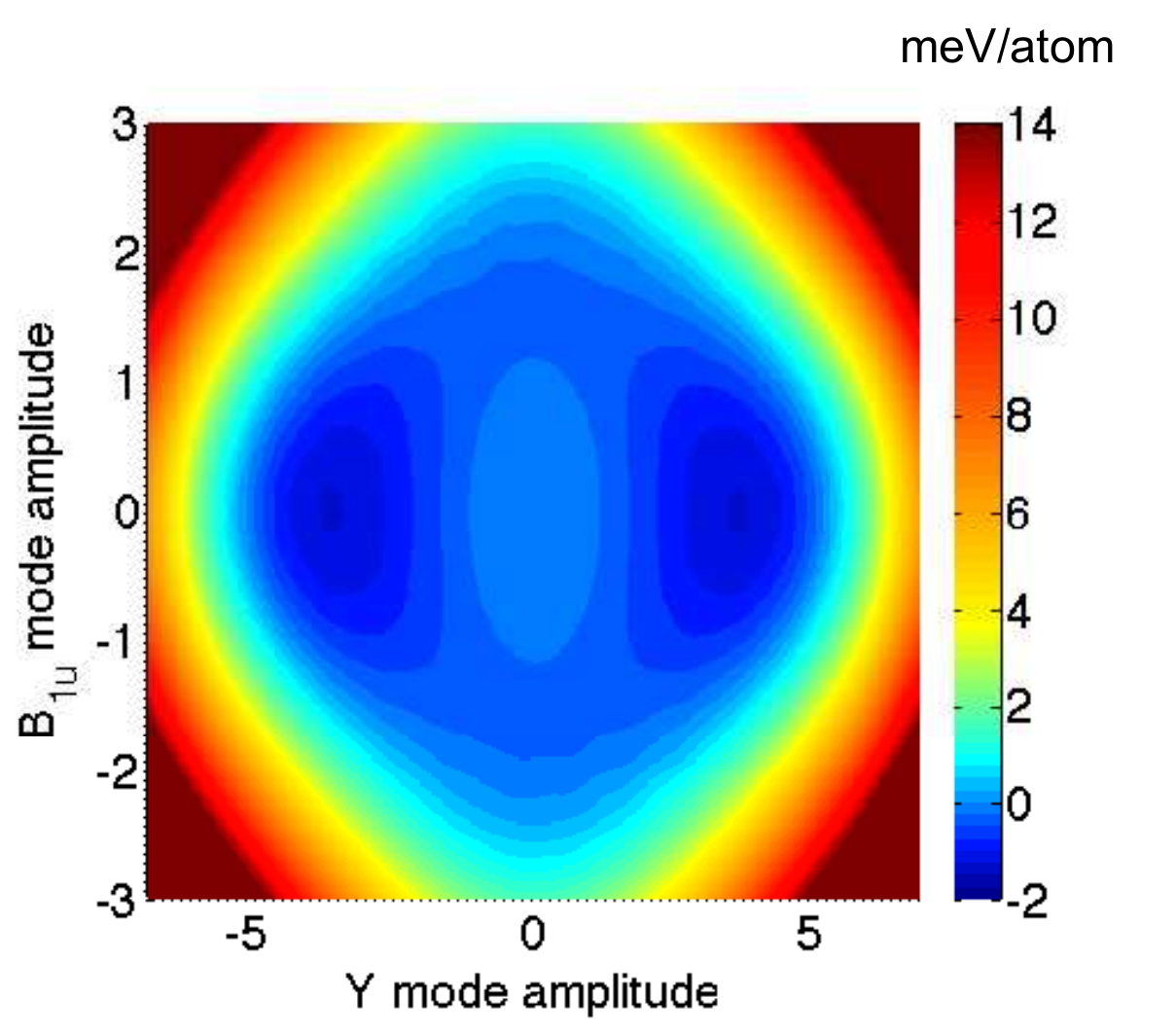}
\caption{\label{fig:YB1u}
(Color online) Frozen phonon energy for
different coupling between $Y$ and $B_{1u}$ mode. 
}
\end{figure*}

The force constants $k_{\alpha \beta}(jl,j'l')$  are defined in a conventional
way:~\cite{Dove,Bruesch}
\begin{equation}
	k_{\alpha \beta}(jl,j'l') = \frac{\partial^2 V}{\partial r_{\alpha}(jl)
	\partial r_{\beta}(j'l')}  = -\frac{\partial F_{\beta}(j'l')}{\partial
		r_{\alpha}(jl)}
\end{equation}
where $\alpha$, $\beta$ are the Cartesian indices, $j,j'$ are the indices of atoms in
a unit cell, and $l,l'$ are the indices of unit cells, $V$ is the potential
energy as a function of atomic position $\mathbf{r}(jl)$. 
According to this definition, a negative force constant indicates that the force
on atom $(j’l’)$  has the same direction as the displacement of atom $(jl)$.  For
example, if the displacement of atom $(jl)$ is along $+\alpha$ direction, the negative
means its displacement induces force on atom $(j’l’)$ along $+\beta$ direction, and vice
versa. 

\begin{figure*}
\includegraphics[width=6in]{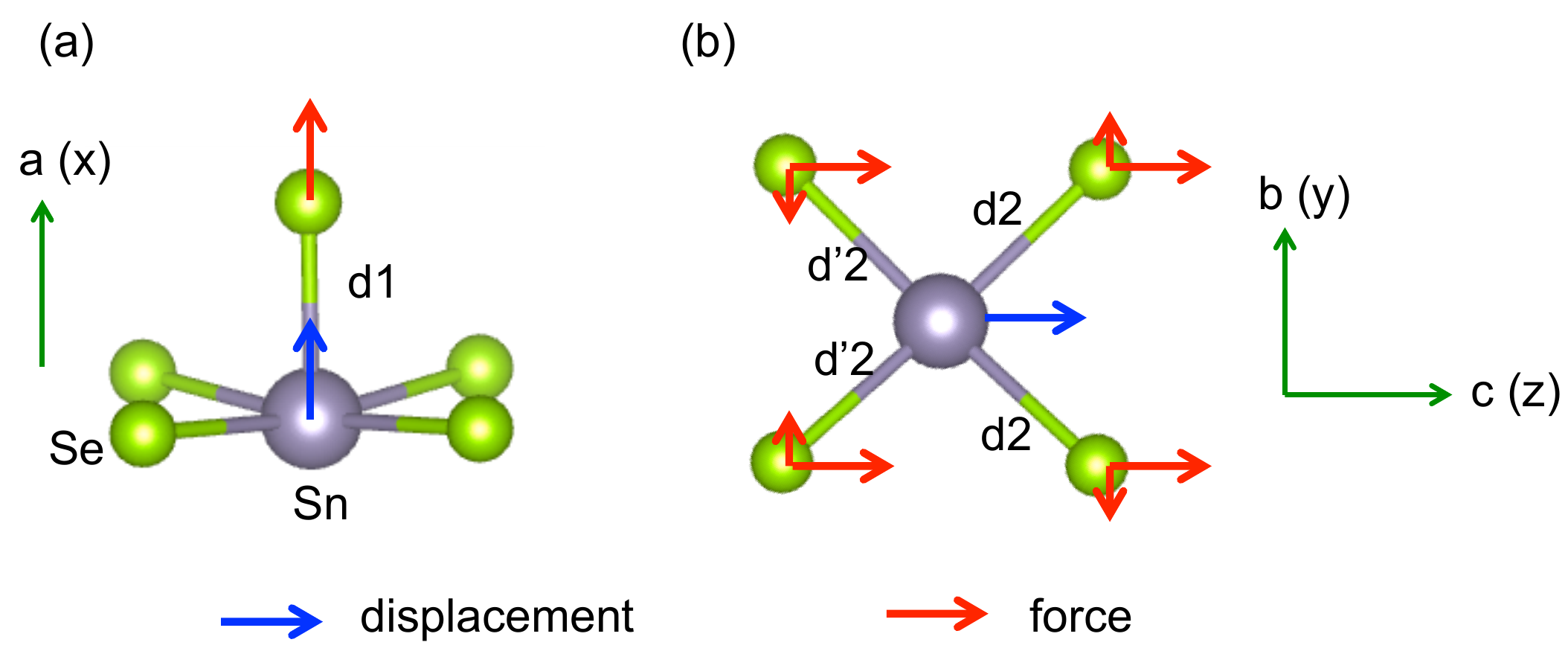}
\caption{\label{fig:FC}
(Color online)  Schematic illustration of negative force constant $k_{xx}$ for $d_1$ bond
(a) and $k_{zz}$ for $d_2$ bonds (b) with Sn displacement (blue arrow). The
negative force constant (listed in Table I in main text) indicates the force on
Se atoms (red arrow) has the same direction as Sn displacement. Sn displacement
along $+c$ also induces force on surrounding Se atoms along $b$ direction
$k_{zy}$, which tends to rotate the bonds, as (b) shows. This simple picture
actually agrees well with the structure evolution from Cmcm to Pnma: $d_2$ bonds
become short and the bond angle become larger, while $d'_2$ bonds become longer
($d_3$ bonds in Pnma phase) and the bond angle become smaller.
} 
\end{figure*}

In main text Table I, the negative force constant
$k_{xx}$ for the bond $d_1$ means that when Sn moves “into” this bond direction (blue
arrow along $+x$ in Fig.~\ref{fig:FC}a), the induced force on Se (red arrow) is
oriented along along $+x$. This is physically reasonable, since the Sn
displacement shortens the $d_1$, which will repel the Se1 atom. For $d_2$ bonds, a
similar schematic explains the signs just as simply. Fig.~\ref{fig:FC}b 
shows a Sn displacement along $c (+z)$ direction and the forces on surrounding Se
atoms. The negative force constant $k_{zz}$  for $d_2$ means that the induced forces on
Se2 is along the same direction $(+z)$ as the Sn displacement. This displacement
also induces forces on neighboring Se atoms along the $b$ direction ($k_{zy}$) , which
tends to rotate the bonds. This simple picture actually agrees well with the
structure evolution from Cmcm to Pnma: $d_2$ bonds become short and the bond angle
become larger, while $d’_2$ bonds become longer ($d_3$ bonds in Pnma phase) and the
bond angle become smaller.

\begin{figure*}
\includegraphics[width=4.0in]{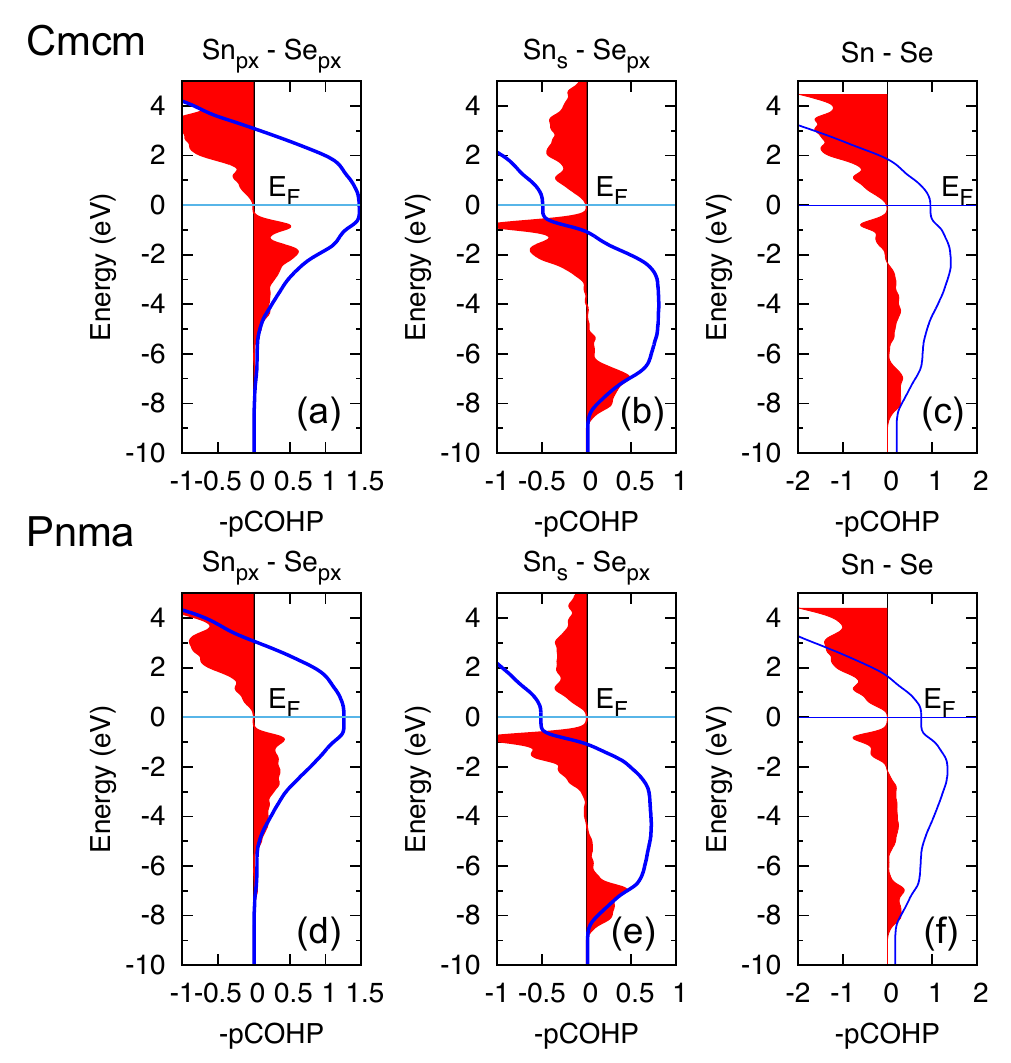}
\caption{\label{fig:COHP-d1}
(Color online)  COHP between $d_1$. Top panel: axial Sn-Se $d_1$ bonding in Cmcm phase;
bottom panel: axial Sn-Se $d_1$ bonding in Pnma phase;  
(a,d) bonding interaction between axial orbitals Sn-5\px\ and
Se-4\px; (b,e) the bonding and anti-bonding interaction between lone
pairs Sn-5$s$ and  Se-4\px\ orbitals; (c,f) the total bonding
interaction between Sn and Se atoms. Red filled curves are the $-$pCOHP and the
blue curves are the $-$IpCOHP.
}
\end{figure*}

\begin{figure*}
\includegraphics[width=6in]{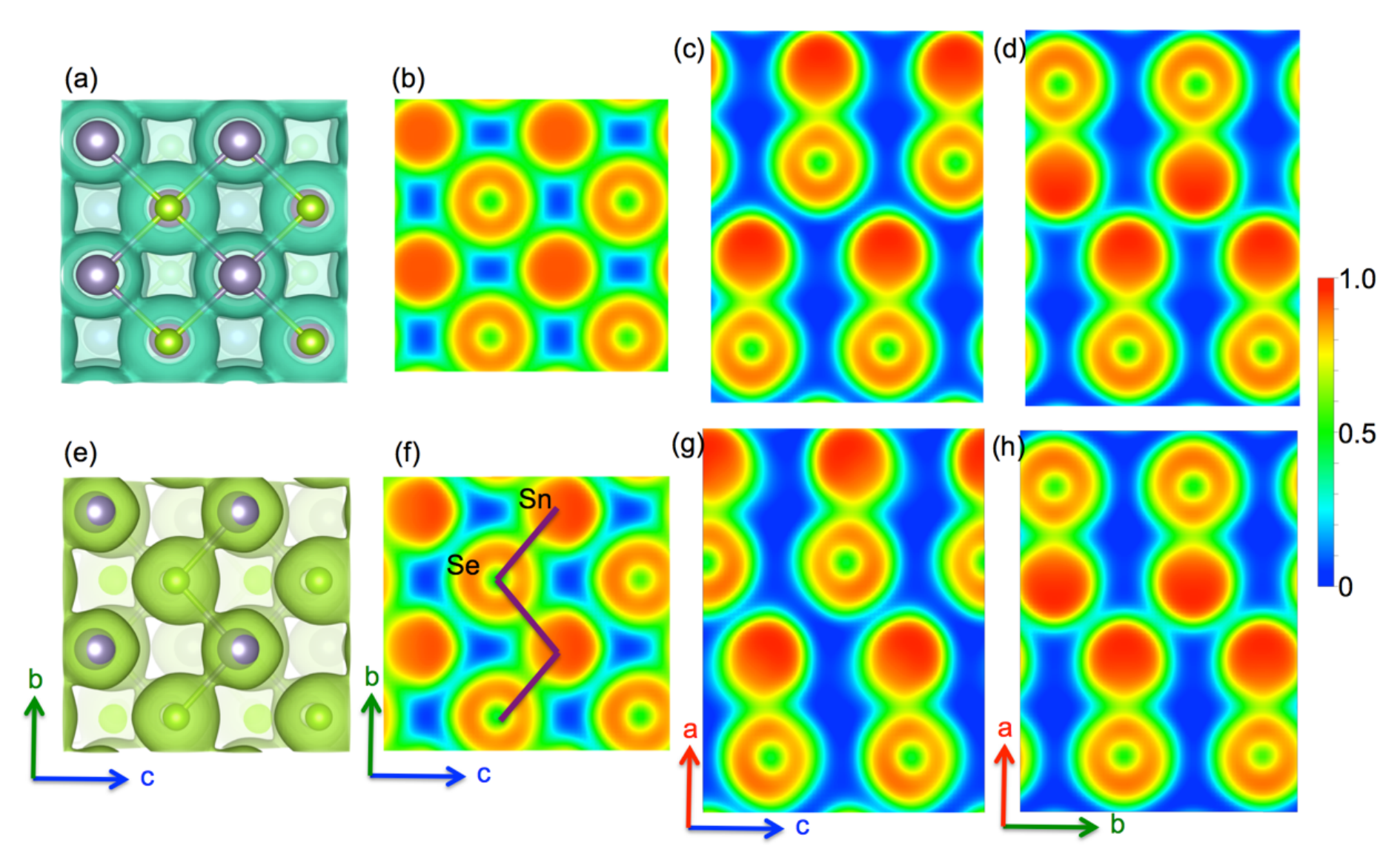}
\caption{\label{fig:ELF}
(Color online) Electron localization functions (ELF)
for Cmcm (top panels) and Pnma (bottom panels) phases. 
(a,e) ELF isosurface in $b-c$ plane, ELF=0.4 for (a) and ELF=0.6 for (e), Sn
atoms are in grey and Se atoms in green;
(b,f) is ELF in $b-c$ plane, (e,g) in $a-c$ plane and (d,h) in $a-b$ plane. 
(c,d,g,h) shows the asymetric large ELF around
Sn along $a$, indicating stereochemically active electron lone pairs. (a, b) show
ELF is symetric around Sn in $b-c$ plane but it becomes active in Pnma phase in
(e,f). The positions of Sn and Se atoms and \dtwo\ bonds (brown thick lines) are
shown in (f). The red end of the color scale corresponds to high electron localization and the blue end
indicates zero zero localization. 
} 
\end{figure*}

\begin{figure*}
\includegraphics[width=4in]{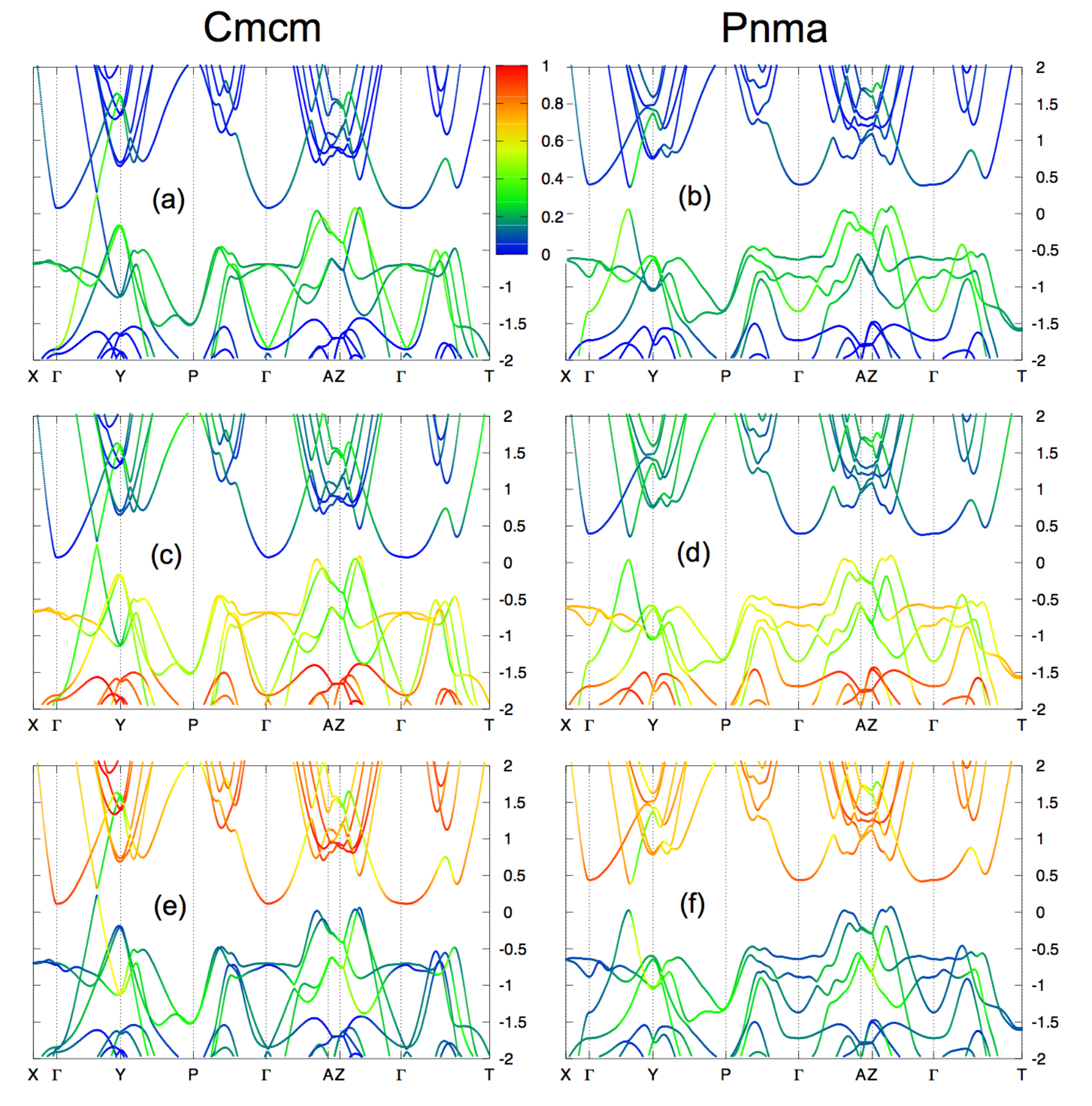}
\caption{\label{fig:eband}
(Color online) Wannier-interpolated electronic band structure for Cmcm phase
(left column) and Pnma phase (right column). Fermi energy is shifted to zero.
(a,b) Sn-s orbital weighted; (c,d) Se-4$p$ weighted; (e,f) Sn-5$p$ weighted. Color
coding indicates the weight of each orbitals. Wannier-interpolated electronic
band structure was calculated by using Wannier90 package.~\cite{wannier90}
}
\end{figure*}
\begin{figure*}
\includegraphics[width=6.5in]{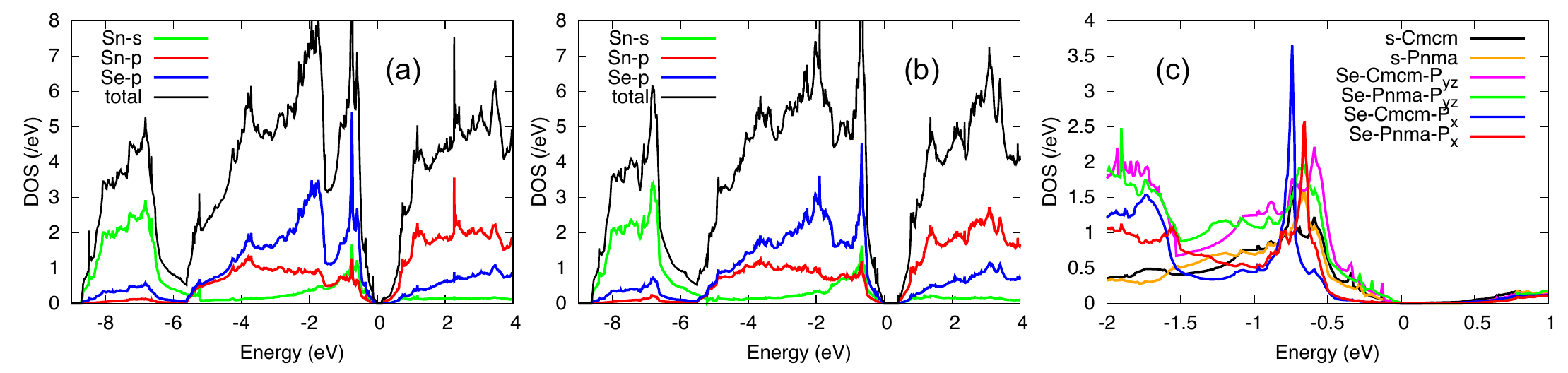}
\caption{\label{fig:dos}
(Color online) Electronic density of state (DOS) for Cmcm (a) and Pnma (b)
phases. (c) is the projected DOS for critical orbitals in both Cmcm and Pnma
near Fermi energy which is shifted to zero. Projected Sn $s$ and Se $p_{yz}$ orbitals
shift towards lower energy side for Pnma phase, but Se $p_x$ shifts towards high
energy level, agrees with COHP analysis.}
\end{figure*}


\end{document}